\documentclass[aps,prb,twocolumn,showpacs,showkeys,amsmath,amssymb]{revtex4}

\usepackage{graphicx}
\usepackage{dcolumn}
\usepackage{bm}
\usepackage{subfigure}

\begin{document}

\preprint{PRB/draft}

\title{Ultrasound attenuation and a $P$-$B$-$T$ phase diagram of superfluid $^3$He in 98\% aerogel}
\author{B. H. Moon}
\author{N. Masuhara}%
\author{P. Bhupathi}
\author{M. Gonzalez}
\author{M.W. Meisel}
\author{Y. Lee}
 \email{yoonslee@phys.ufl.edu}
\affiliation{%
National High Magnetic Field Laboratory and Department of Physics, University of Florida, Gainesville, Florida 32611-8440, USA}%

\author{N. Mulders}
\affiliation{
Department of Physics and Astronomy, University of Delaware, Newark, Delaware 19716, USA}%

\date{\today}

\begin{abstract}
Longitudinal sound attenuation measurements in superfluid $^3$He in 98\% aerogel were conducted at pressures between 14 and 33 bar and in magnetic fields up to 0.444 T. The temperature dependence of the ultrasound attenuation in the $A$-like phase was determined for the entire superfluid region by exploiting the field induced meta-stable $A$-like phase at the highest field. In lower fields, the $A-B$ transition in aerogel was identified by a smooth jump in attenuation on both cooling and warming. Based on the transitions observed on warming, a phase diagram as a function of pressure ($P$), temperature ($T$) and magnetic field ($B$) is constructed. We find that the $A-B$ phase boundary in aerogel recedes in a drastically different manner than in bulk in response to an increasing magnetic field. The implications of the observed phase diagram are discussed.
\end{abstract}

\maketitle
\section{Introduction}
The influence of high porosity aerogel as quenched disorder has been studied in various systems such as liquid $^4$He,\cite{chan} $^3$He-$^4$He mixture,\cite{kim,McRae} $^3$He,\cite{porto,sprague} and liquid crystals.\cite{bellini}  The effect of aerogel on superfluid $^3$He is exceptionally interesting because it is a {\it p-wave triplet} anisotropic superfluid possessing continuous symmetry.  Since the discovery of superfluiditiy of $^3$He in high porosity aerogel,\cite{porto,sprague} more than a decade of theoretical and experimental efforts have been invested to understand this system and have revealed many interesting phenomena.  The fragile nature of {\it p-wave} pairing against impurity-scattering was immediately recognized by the significant depression of superfluid transition, \cite{porto,sprague,matsumoto} and the theoretical descriptions based on various isotropic impurity-scattering models have provided a successful account for the observed behavior. \cite{thunne,hanninen,sauls03} A wide variety of experimental evidence reflecting the role of aerogel as an effective pair-breaking agent are now well documented.\cite{halperin}

For the past few years, attention has been shifted to understand phenomena related to an energy scale smaller than the condensation energy. For example, the relative stability among possible superfluid phases, specifically the transition between two superfluid phases observed in this system, the $A$-like and the $B$-like phases, has been investigated.  In the absence of a magnetic field, the supercooled $A$-like phase appears at all pressures studied, even below the bulk polycritical point (PCP),\cite{gervais,nazaretski,choithesis} while only a very narrow region where the two phases coexist was identified on warming.\cite{vicente} In the presence of low magnetic fields, the $B$-like to $A$-like transition was observed, on warming, to follow a quadratic field dependence,\cite{brussard,gervais,baumgardner} which is reminiscent of the bulk $A-B$ transition, $1-T_{AB}/T_{c}=g(\beta)(B/B_{c})^{2}$, where $T_{AB}$ and $T_{c}$ are the $A-B$ transition and the superfluid transition temperatures, respectively, and $g(\beta)$ is a strong-coupling parameter that is a function of $\beta$ parameters of the Ginzburg-Landau free energy (see Sec. III). However, the systematic field and pressure dependence study by Gervais {\it et al.\,}\cite{gervais} found a monotonic increase in $g(\beta)$ with pressure without showing any anomalies.  This observation raised a question on the position or the existence of the PCP in aerogel. It is important to emphasize that the $A$ and the $B$ phases of bulk $^3$He are highly competing phases separated by first-order transition with a minute-free-energy difference and have identical intrinsic superfluid transition temperatures.  These properties are at the heart of many intriguing phenomena showing subtle modifications of the $A-B$ transition in the presence of weak external perturbations such as a magnetic field.  Therefore, it is reasonable to expect that the presence of impurities or disorder will have a similar influence on the $A-B$ transition.

In 1996, Volovik\cite{volovik96} discussed the significance of the quenched random anisotropic disorder presented by the strand-like aerogel structure and its interaction with the anisotropic order parameter. This coupling is thought to be particularly important in the $A$-phase, where the order parameter is doubly anisotropic in the sense that the rotational symmetries in spin and orbital space are broken separately. Vicente {\it et al.\,}\cite{vicente} argued that the aerogel strands generated orbital fields emulating the role of a magnetic field, thereby giving rise to similar profound effects on the $A$-like to $B$-like transition. They further suggested the use of uniaxially deformed aerogel to amplify and to systematically investigate the effect of the anisotropic disorder.\cite{vicente}  A series of calculations by Aoyama and Ikeda\cite{aoyama,aoyama3} are consonant with these ideas and predict a widened $A$-like phase region in a uniaxially deformed aerogel, the appearance of a novel superfluid phase in uniaxially stretched aerogel, and a change in the PCP location in the phase diagram.

Unlike the $B$-like phase, the clear identification of the $A$-like phase in aerogel has not been made.  However, some of the recent NMR measurements using uniaxially deformed aerogels\cite{kunimatsu,elbs} provide compelling evidence that the $A$-like phase possesses the $ABM$ pairing symmetry, albeit with unusual textural configurations.  The free-energy calculation by Ikeda and Aoyama\cite{ikeda} also found the disordered $ABM$ phase as the most stable among the various plausible pairing states, such as the Imry-Ma,\cite{volovik06} the planar, and the robust\cite{fomin} phases.  Furthermore, the third superfluid phase observed in 98\% aerogel in the presence of high magnetic fields\cite{choia1} fortifies this identification.  Therefore, we will continue our discussion with the assumption that the $A$-like phase observed at least in 98\% aerogel has the same pairing symmetry as the bulk $A$-phase.

With this notion, we conducted longitudinal ultrasound attenuation measurements in the superfluid phases of $^3$He in 98\% porosity silica aerogel. Our measurements were performed in the presence of magnetic fields, 0 to 0.444~T, and at various sample pressures ranging from 14 to 33~bar. At the highest field, the existence of the meta-stable $A$-like phase persisted to the lowest temperatures, thereby allowing the sound attenuation in the $A$-like phase to be measured over the entire range of the temperatures studied. In lower magnetic fields, we were able to identify the transitions between the two phases on cooling and warming, and herein, a $P$-$B$-$T$ phase diagram of this system is presented.

\section{Experiment}
The presence of the compliant aerogel complicates the sound propagation because the sound modes of the liquid $^3$He and the aerogel matrix are effectively coupled.\cite{mckenna}  As a result, two longitudinal sound modes emerge in this composite medium: one with the speed of sound close to, but slightly lower than, that of the liquid (fast mode) and the other with a significantly lower speed of sound (slow mode).\cite{golov}   We measured the longitudinal fast sound attenuation in superfluid $^3$He in 98\% aerogel at frequencies between 3.69 and 11.3~MHz. The employment of the multiple frequency excitations turned out to be extremely valuable in this work for the reason described later in this paper.

Two best-matched LiNbO$_3$ transducers (9.6~mm diameter) with fundamental resonances of 1.1~MHz were selected from six transducers tested using a broadband spectrum analyzer and were used as a transmitter and a receiver. The transducers were supported by a MACOR spacer forming a 3.02~mm size acoustic cavity.   Aerogel with 98\% porosity was grown in and around this cavity to ensure optimal acoustic coupling between the aerogel and the transducers. The aerogel grown outside of the cavity was carefully removed, and copper wires were attached to the outer surfaces (electrodes) of the transducers using silver epoxy.  In order to reduce the ringing of the transducers, a thin layer of silver epoxy was applied to the electrode.  A small piece of a cigarette paper with numerous needle holes was placed between each transducer and the cell wall to interrupt back reflections from the wall through the bulk liquid.  The sample cell housing the cavity was placed inside a homemade superconducting solenoid magnet located in the inner vacuum space.  The magnet was thermally anchored to the mixing chamber. We chose the magnetic field, $\vec{B}$, to be perpendicular to the sound wave vector $\vec{q}$, $\vec{B}\bot\vec{q}$, expecting ${\vec{l}}\parallel\vec{q}$ in the $A$-like phase, where $\vec{l}$ indicates the orbital angular momentum of the Cooper pair. The top part of the sample cell forms a diaphragm so the pressure of the cell can be measured capacitively. The variation in the cell pressure during the measurement was around $\pm0.1$~bar. A schematic of the experimental geometry can be found elsewhere.\cite{moon}

A commercial spectrometer, LIBRA/NMRKIT II (Tecmag Inc., Houston, TX) was used to transmit 3 $\mu$s pulses and also to detect the transmitted signals.  Each measurement was obtained by averaging eight transmitter signals produced in a phase alternating pulse sequence.  The level of excitation used in this experiment was set in the range where neither self-heating nor nonlinearity was observed.  In one temperature sweep, the measurements at four pre-determined frequencies were performed in a cyclic manner.  The temperature was monitored by a melting curve thermometer for $T\geq1$~mK and a Pt-NMR thermometer for $T\leq1$~mK.

In spite of our effort to spoil the quality factor of the transducers, sustained ringings were observed and we were unable to resolve echoes following the initial received signal. Consequently, by integrating a portion of the received signal, only the relative attenuation could be determined. The region of integration was carefully chosen not to include any echoes.  Our method produced consistent relative attenuation for various choices of the integration range within the safe window described above. The relative attenuation in reference to the value at the aerogel superfluid transition temperature ($T_{ca}$) was determined by
\begin{equation}
\Delta\alpha=\alpha(T)-\alpha(T_{ca})=-\frac{1}{d}\ln\frac{A(T)}{A(T_{ca})},
\end{equation}
where $d$ is the sound path length and $A(T)$ is the integrated area of the transmitter signal at temperature $T$.

\begin{figure}
\includegraphics[width=3.7in]{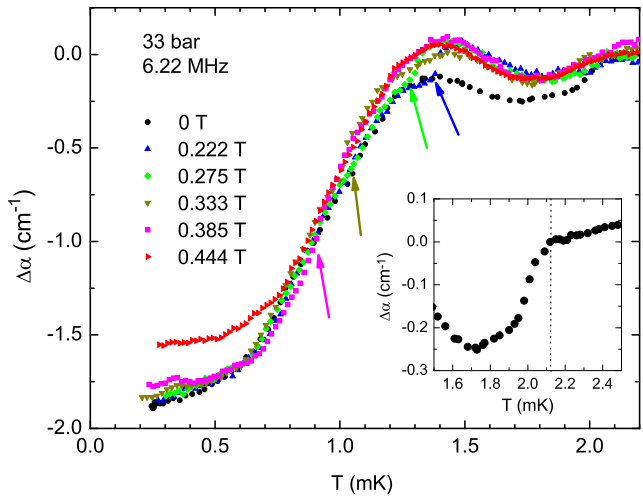}
\caption{\label{Fig.1.} (Color online) Temperature dependences of relative longitudinal sound attenuations using a 6.22~MHz excitation at 33 bar in the presence of various magnetic fields.  All the data were taken on warming after cooling through the $A$-like to $B$-like transition except for $B =$ 0.444~T, where no supercooled transition was observed.  The arrows point the positions where the $B$-like to $A$-like phase transitions occur. Inset: expanded view of zero-field attenuation near the superfluid transition indicated by the vertical line.}
\end{figure}

\begin{figure}
\includegraphics[width=3.7in]{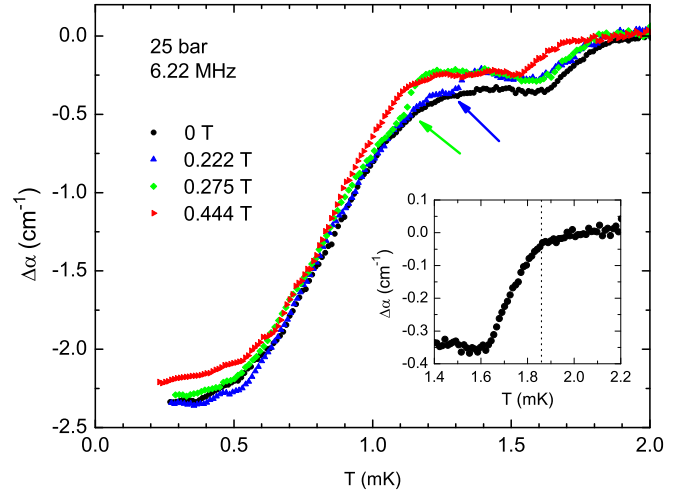}
\caption{\label{Fig.2.} (Color online) Temperature dependences of relative longitudinal sound attenuations using a 6.22~MHz excitation at 25~bar in the presence of various magnetic fields. See the caption of Fig.\,1 for additional details. Inset: expanded view of zero-field attenuation near the superfluid transition indicated by the vertical line.}
\end{figure}

\section{Results and discussion}
\subsection{Longitudinal sound attenuation and the $A-B$ transition in aerogel}

Figures 1 and 2 show the relative ultrasound attenuations obtained at 33 and 25~bar in the presence of magnetic fields ranging from 0~T to 0.444~T, respectively. All the data shown were taken on warming after cooling though the supercooled $A$-like to $B$-like transition at a fixed external magnetic field, except for $B=$~0.444~T, where no supercooled transition was observed down to $\approx$~200~$\mu$K.  Therefore, the warming trace at the highest field should be in the $A$-like phase for the entire temperature range, probably in the meta-stable $A$-like phase in the low temperature region.  The superfluid transition is marked by a slight decrease in attenuation around 2.1~mK for 33~bar (Fig.\,1) and 1.9~mK for 25~bar (Fig.\,2). The zero field attenuation, which essentially represents the $B$-like phase attenuation except for a very narrow region ($\approx$~100~$\mu$K) right below $T_{ca}$, can be directly compared with the absolute attenuation measurements by Choi {\it et al.\,}\cite{choi} performed under almost identical experimental conditions. The features observed in the current experiment, namely, the broad shoulder structure appearing in the range $1.0 < T < 1.5$~mK and the absence of attenuation peak associated with the pair-breaking and the order-parameter collective modes, are consistent with those reported earlier\cite{choi} and also with the calculations by a Hiroshima group.\cite{higashitani}
\begin{figure}
\includegraphics[width=3.7in]{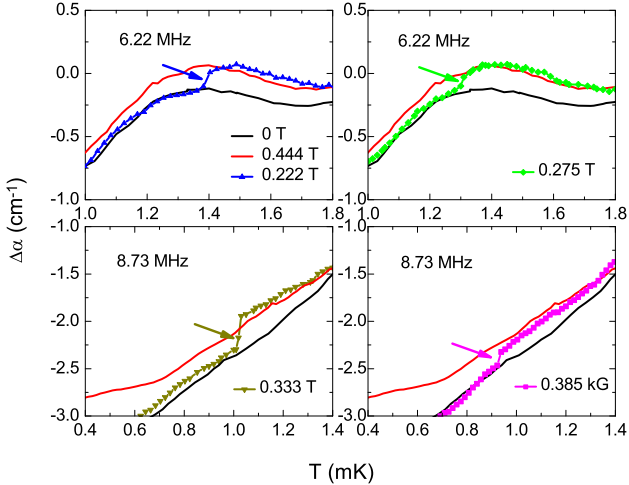}
\caption{\label{Fig.3.} (Color online) The $A-B$ transition features in sound attenuation at 33 bar. The red or upper (black or lower) trace represents the attenuation in the $A$-like ($B$-like) phase.  The top (bottom) panels show the traces taken using 6.22~MHz (8.73~MHz) excitations.  The switching behavior between the two phases is clearly demonstrated for each field as marked by an arrow.}
\end{figure}

\begin{figure}
\includegraphics[width=3.7in]{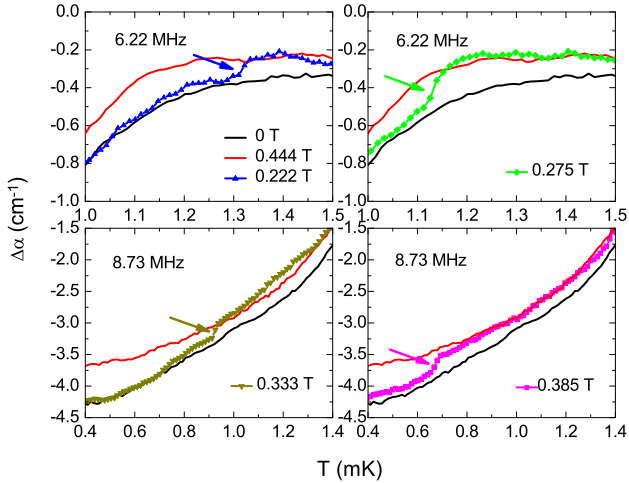}
\caption{\label{Fig.4.} (Color online) The $A-B$ transition features in sound attenuation at 25 bar. See the caption of Fig.\,3 for additional details. }
\end{figure}

\begin{figure}
\includegraphics[width=3.7in]{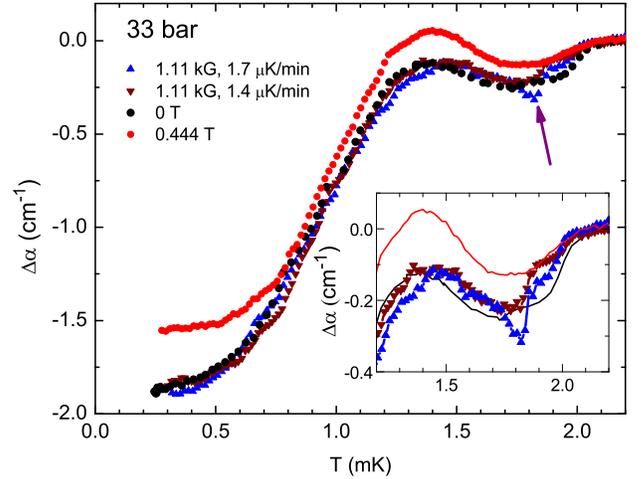}
\caption{\label{Fig.5.} (Color online) Temperature dependence of attenuation at 33~bar using 6.22~MHz excitation at two different warming rates.  The attenuation in the $B$-like ($B =$~0) and the $A$-like ($B=$~0.444~T) phases are already shown in Fig.\,1. For $B =~$0.111~T, the attenuation was measured with two warming rates of 1.4~$\mu$K (inverted triangles) and 1.7~$\mu$K (triangles). Inset: magnified view of the region of the $A-B$ transition in aerogel.}
\end{figure}

Establishing the attenuation in the $A$-like ($B=$~0.444~T) and the $B$-like ($B=$~0) phases for the entire temperature range in the superfluid, one can envision a transition between the two phases at any intermediate field where a switching from one trace to another occurs. It is expected that the attenuation in the $A$-like phase is higher than in the $B$-like phase under the assumption that it is the $ABM$ state, since the sound presumably propagates along the node direction in our experimental configuration. However, unlike in the bulk, the difference in attenuation between the $A$-like and the $B$-like phases is much smaller and subtle  because of the absence of the order-parameter collective modes, which are the fingerprints of specific pairing symmetry, and the presence of the impurity states residing in the gap.  One can see the subtle difference in the attenuation between two phases in Figs.\,1 and 2.  At all temperatures, the attenuation in the $A$-like phase is slightly larger than in the $B$-like phase, while the largest difference is observed in the zero temperature limit.  For this reason, the acoustic signature of the $A-B$ transition in aerogel is not as clear as in the bulk.  Despite this small difference in attenuation, the $B$-like to $A$-like transition features are noticeable in most of the cases (indicated by the arrows in Figs.\,1 and 2).  However, in the temperature region where two phases show almost identical attenuation, as in $0.7<T<1.0$~mK or very close to $T_{ca}$, the transition feature is rather vague. When this situation arose, the transition temperature $T_{ABa}$ was determined from the attenuation measurements conducted at other frequencies. The magnified views of the $A-B$ transition features are shown in Figs.\,3 and 4 for 33 bar and 25 bar, respectively.  For each field, the switching behavior between the $A$-like (red, upper trace) and the $B$-like (black, lower trace) phases is unmistakably demonstrated in these figures. While the transitions at $B=0.333$ and 0.385~T for 33 bar at 6.22~MHz (Fig.\,1) are not clear, the transitions at 8.73~MHz are much more evident in Fig.\,3.  This phenomenon is due to the non-trivial frequency dependencies of the attenuation observed in aerogel.  The details of this subject are beyond the scope of this paper and will be reported in a separate publication. 

\begin{figure}
\includegraphics[width=3.8in]{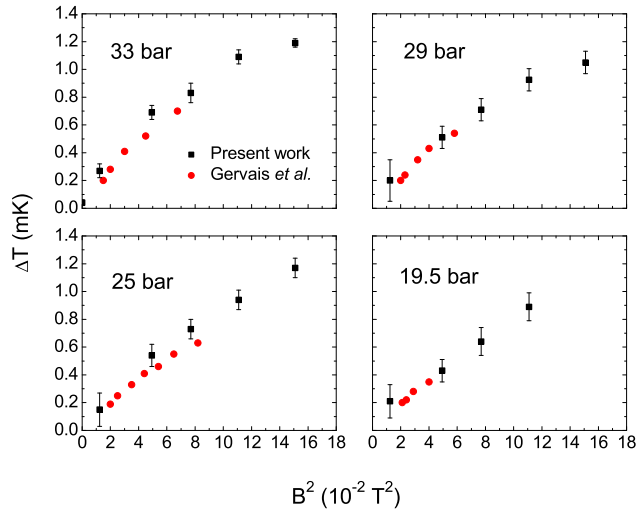}
\caption{\label{Fig.6.}(Color online) Magnetic field dependence of the width of the $A$-like phase, $\Delta T=T_{ca}-T_{ABa}$. For comparison, our results are plotted along with those from Gervais {\it et al.\,} (solid circles)\cite{gervais}.  The data points from Gervais {\it et al.\,} were taken at the slightly different pressures of 33.4, 28, 25, and 20~bar, respectively.}
\end{figure}

The lowest finite magnetic field used in this experiment was 0.111~T, and two attenuation measurements performed in this field at 33~bar are shown in Fig.\,5.  These data were collected with two different warming rates of 1.4~$\mu$K/min (inverted triangles) and 1.7~$\mu$K/min (regular triangles).  Both measurements produced the same transition temperature despite the difference in the warming rate by about 20\%. The small differences between the two traces arise from the background drift associated with the $^4$He bath level and room-temperature variation.

In Fig.\,6, the widths of the $A$-like phase, $\Delta T=T_{ca}-T_{ABa}$, as a function of $B^2$, along with the results obtained in the low-field region by Gervais {\it et al.\,}, are plotted.  Within the Ginzburg-Landau (G-L) limit, we can perform analysis that is similar to work used to describe the bulk liquid.\cite{tang} Specifically, the suppression of the $B$-like phase in finite magnetic fields can be written as
\begin{equation}
1-T_{ABa}(T)/T_{ca}=g(\beta)(B/B_{c})^{2}+\mathcal{O}(B/B_{c})^{4}.
\end{equation}
Here, $B_{c}$ represents a characteristic field scale directly related to the transition temperature, namely,
\begin{equation}
B_{c}=\sqrt{\frac{8 \pi^2}{7\zeta(3)}}\frac{k_{B}T_{ca}}{\gamma\hbar}(1+F^{a}_{0}),
\end{equation}
where $k_B$, $\gamma, \zeta(x)$, and $F^{a}_{0}$ are the Boltzmann constant, the gyromagnetic ratio for a $^3$He nuclei, the Riemann zeta function, and a Fermi-liquid parameter, respectively. In addition, the strong-coupling parameter $g(\beta)$ is a function of the pressure-dependent $\beta$-parameters, the coefficients of the quartic terms in the G-L free-energy expansion,\cite{fetter} and can be written as
\begin{multline}
g(\beta)= \frac{\beta_{245}}{2(2\beta_{345}-3\beta_{13})} \times \\
\left(1+\sqrt{\frac{(3\beta_{13}+\beta_{345})(2\beta_{13}-\beta_{345})}{\beta_{245}\beta_{345}}}\right),
\end{multline}
where $\beta_{ijk}=\beta_{i}+\beta_{j}+\beta_{k}$.  In the weak coupling limit, $g(\beta)\rightarrow$~1, and the strong-coupling effects cause it to increase.

\begin{figure}
\includegraphics[width=3.7in]{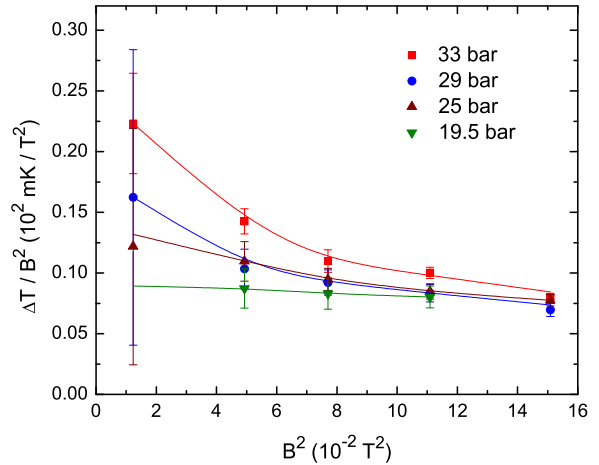}
\caption{\label{Fig.7.} (Color online) Magnetic field dependence of the width of the $A$-like phase scaled by $B^2$. The quadratic coefficient, $\textit{g}(\beta)$ is determined by the intersection of the each curve with the $B=$~0 axis, Eq.\,(2).}
\end{figure}

In order to illuminate the overall field dependence, the data presented in Fig.\,6 are recasted as $\Delta T/B^{2}$ in Fig.\,7.  As noted by Tang {\it et al.\,},\cite{tang} one of the advantages of this plot is that the intersection of the curve with the $B=0$ axis gives the strong-coupling parameter, $g(\beta)$, and the slope of the curve is related to the coefficient of the higher-order correction, as can be seen in Eq.\,(2).  Our $g(\beta)$ values extracted by extrapolating to zero field are shown in Fig.\,8.  In the same figure, $g(\beta)$ of the bulk by Tang {\it et al.\,} (open circles) and of 98\% aerogel by Gervais {\it et al.\,} (solid cricles) are included for comparison.  Additionally, we reproduced the theoretical calculation\cite{gervais} based on the homogeneous scattering model (HSM)\cite{thunne} with the rescaled strong-coupling corrections by the factor of $T_{ca}/T_c$ for two different mean-free path values of $\ell =$~150 (dotted-dashed line) and 200~nm (dashed line).  Although our $g(\beta)$ value at 19.5~bar is in good agreement with that of Gervais {\it et al.\,}, the discrepancy between the two sets of data becomes larger at higher pressures.  However, $g(\beta)$ in aerogel from both measurements is substantially smaller than that of the bulk value at the corresponding pressure. For the bulk, $g(\beta)$ grows quickly and approaches the PCP as predicted by the G-L theory.  However, no such behavior is seen in aerogel.  Although the error bars in our data are rather large, our results lie between the two theoretical curves.
It is also interesting to observe that the sign of the quartic correction is negative at higher pressures and seems to change its sign at $P\approx 19.5$~bar (see Fig.\,7), which needs to be compared with the bulk case where the sign crossover occurs at $P\approx 6.7$~bar.\cite{tang}  Based on these observations, one could argue that the presence of aerogel reduces the strong coupling effects and, in effect, the phase diagram of this system is shifted up in pressure.

\begin{figure}
\includegraphics[width=3.7in]{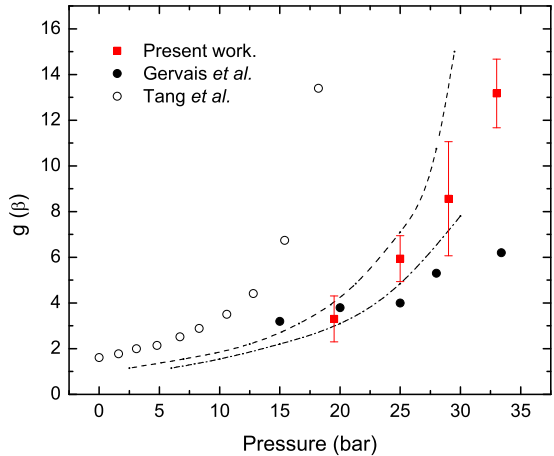}
\caption{\label{Fig.8.} (Color online) Pressure dependence of $g(\beta)$. The present data (solid squares) are shown   with the data by Gervais {\it et al.\,}(solid circles)\cite{gervais} for aerogel and by Tang {\it et al.\,}(open circles)\cite{tang}  for the bulk liquid. The dashed and doted-dashed lines are from HSM with the transport mean-free path, $\ell=$~200  and 150~nm, respectively (see Ref.\,12 for details).}
\end{figure}

\subsection{$A-B$ transition in aerogel by isothermal field sweeps}

The $A-B$ transition can be induced through an isothermal field sweep (IFS).  Although it is a time-consuming process, an IFS offers an independent way of determining this phase transition and is especially valuable in the region where the slope of the transition curve in the $T-P$ phase diagram becomes small.  During an IFS in either the up or down direction, heating was observed due to the eddy currents in the silver cell body. To alleviate this problem, we slowly demagnetized the main magnet of the nuclear demagnetization stage during a field sweep (typically $\approx 14~\mu$ T/min). This passive procedure limited the temperature variation during an IFS to $\approx$~50~$\mu$K.

In Fig.\,9, the magnitudes of the integrated acoustic signals taken at four different frequencies during an isothermal field sweep at 25~bar and 0.3~mK are displayed.  The temperature variation during this process is also shown in the same figure.  The sample was cooled from the normal fluid in the presence of a magnetic field of 0.444~T to $\approx$~0.3~mK.  After establishing equilibrium, the magnetic field was slowly reduced at the rate of 0.4~mT/min\cite{note1} to go through the $A$-like to $B$-like transition.  Therefore, the $B$-like phase was supposed to be induced through a primary nucleation, and this case is the only instance of a primary nucleation transition observed by IFS in our work.  For the entire sweep process, the temperature remained within $\approx$~30~$\mu$K around 0.27~mK.  The smooth change in magnitudes at all frequencies can be observed from $\approx$~0.43 to 0.4~T, indicating the transition from the $A$-like to $B$-like phase.  The difference in the magnitude of the acoustic signal between two phases matches well with the attenuation difference determined from the temperature sweep measurements shown in Fig.\,2.

For $B\lesssim$~0.4~T (in the $B$-like phase), the attenuation exhibits a weak-field dependence, most notably at 11.3~MHz.  This behavior can not be simply attributed to the temperature variations during the field sweep because the attenuation shows a very weak temperature dependence around 0.3~mK (see Figs.\,1 and 2).  One can speculate that this variation in attenuation might be related to the progressive distortion of the gap induced by magnetic field, as the isotropic $BW$ state evolves through the distorted $BW$ state to the planar state and eventually to the $ABM$ phase with the node along the sound propagation direction.\cite{Teowrdt} The increase (decrease) in the magnitude (attenuation) in the low-field region could be due to the enhancement (reduction) in the component of the gap perpendicular (parallel) to the magnetic field.  In the $A$-like phase at the highest field, the sound propagates in the node direction, resulting in a higher attenuation.

\begin{figure}
\includegraphics[width=3.7in, height=3in]{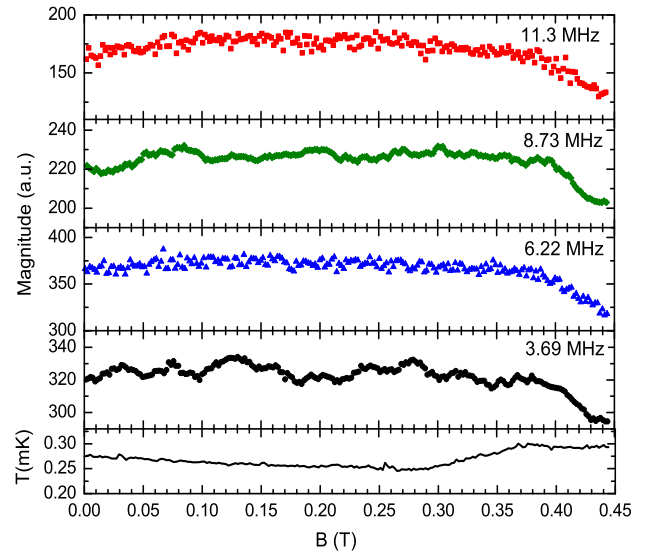}
\caption{\label{Fig.9.} (Color online) Results of the IFS at 0.3~mK and $P=$~25~bar. The magnitudes of the integrated acoustic signals, $A(T)$, measured using four different excitation frequencies are displayed as a function of magnetic field.  The temperature variation during the IFS is also shown in the bottom panel.}
\end{figure}

Several additional IFS studies were conducted at various combinations of pressure and temperature, where the sample was cooled from the normal state at a fixed field to a temperature in the $B$-like phase via the superfluid and the supercooled $A$-like to $B$-like transitions. Then, the magnetic field was ramped up through the $B$-like to $A$-like transition and decreased again back through the transition, if necessary. Figure 10 shows the IFS results at 14 bar and $T\approx 0.27$~mK. The phase transition occurs over a rather broad range of field ($\Delta B\approx 50$~mT), but no appreciable hysteresis was observed.  The results of two other IFS studies at 29~bar ($T\approx$~0.86 and 1.38~mK) are shown in Figs.\,11 and 12.  For $T\approx 0.86$~mK (Fig.\,11), the transition can only be identified in the 3.69~MHz measurements ($\Delta B\approx$~20~mT).

\begin{figure}
\includegraphics[width=3.7in]{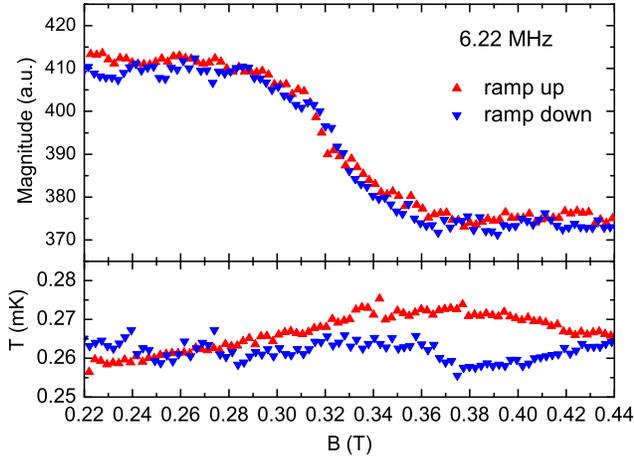}
\caption{\label{Fig.10.} (Color online) Results of the isothermal field sweep at 14 bar.}
\end{figure}

Brussaard {\it et al.\,}\cite{brussard} observed hysteretic behavior in the field driven $A-B$ transition in their measurements at $T \approx$~0.335~mK and $P=$~7.4~bar using an oscillating aerogel sample attached to a vibrating wire.  The magnetic field sweep was performed in the presence of a field gradient in which a single A-B phase boundary was moving through the sample during the process. They proposed the pinning of the $A-B$ phase boundary by the aerogel strands as a mechanism for the observed hysteresis.  Furthermore, based on this scenario, they made an argument that the $A-B$ transitions determined by a conventional temperature sweep method, specifically those by Gervais {\it et al.\,}, might not provide reliable thermodynamic transition points due to supecooling and superwarming caused by the pinning, suggesting the finite width of the transition is an evidence of the existence of a range of pinning potential strengths.\cite{fishercomment}  We would like to point out that the experiments by Gervais {\it et al.\,} and by us were performed without a designed field gradient.  In this case, it is also plausible that the random disorder presented by aerogel, more specifically anisotropic disorder, could cause the broadening of the transition.\cite{imry,vicente}  The effect of rounding by disorder is also apparent in the superfluid transition, which is a second-order transition and does not involve an interfacial boundary.  Imry and Wortis\cite{imry} have made a heuristic argument about the influence of random impurities on a first-order transition.  They predicted various degrees of rounding in the transition due to fluctuations (inhomogeneities) of the random microscopic impurities through the simple generalization of the Harris criterion\cite{harris} valid for second-order transition.  It is worth noting that the Lancaster group also reported a similar degree of hysteresis in field (approximately millitesla) in the bulk $A-B$ transition induced by a similar method.\cite{bartkowiak}  The field sweep performed at 29 bar around 0.86~mK in Fig.\,11 seems to show a glimpse of hysteresis in the 3.69~MHz data.  However, we acknowledge that hysteresis at the level of millitesla can not be resolved from our measurements, and the width of the transition is certainly larger than any hysteresis that might exist.

\begin{figure}
\includegraphics[width=3.7in]{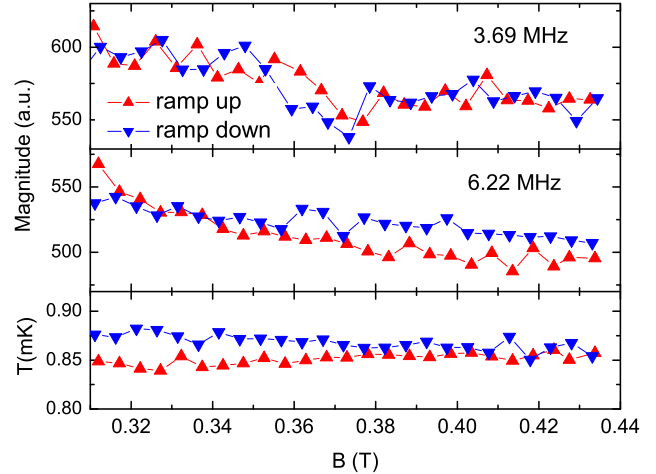}
\caption{\label{Fig.11.} (Color online) Results of the isothermal field sweep at 29~bar and $T \approx 0.86$~mK.}
\end{figure}

\subsection{Phase diagram}
The $A-B$ transitions in aerogel identified by the temperature sweep at constant field (TSCF) and the IFS are plotted in the $P$-$T$ phase diagram in Fig.\,13. For both methods, the mid-point of the transition in $T$ or $B$ was chosen as the transition point and the actual width of the transition is represented by the error bar.  The width in $B$ is translated into the temperature width using the measured field dependence of the $A-B$ transition in aerogel (see Figs.\,1 and 2).   The transition points determined by the two different methods exhibit self-consistency within the resolution of our measurements.  For example, the IFS transition point at 14~bar was observed at 0.33~T and lies on the extension of the TSCF measurements at 0.333~T, and the 0.37~T IFS point at 29 bar is right on the line for 0.385~T from the TSCF.  We could not have obtained the IFS point at 0.421~T and 25~bar by the conventional TSCF at this field.

\begin{figure}
\includegraphics[width=3.7in]{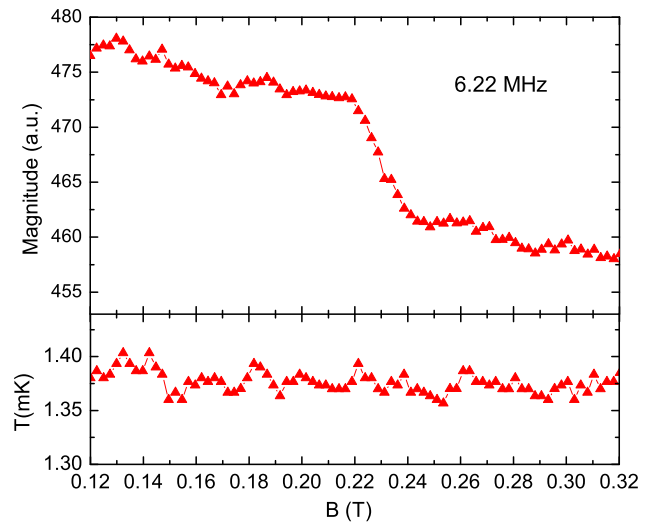}
\caption{\label{Fig.12.} (Color online) Results of the isothermal field sweep (ramp up only) at 29~bar and $T \approx 1.38$~mK.}
\end{figure}

The emerging phase diagram, Fig.\,13, from our measurements unambiguously reveals that the $A-B$ phase boundary in 98\% aerogel recedes toward the melting pressure and zero-temperature corner in response to the increasing field. This tendency is robust even when allowing for the possibility of superwarming, which might shift the transition temperature down.  This phase diagram is in drastic contrast to that of the bulk.\cite{hahn} First, the slope of the constant-field phase boundary is positive in aerogel but negative in bulk for most  of the corresponding pressure range.  Second, the phase boundary in the bulk recedes toward $P\approx$~19~bar, which is in close proximity to the bulk PCP, rather than toward the meting pressure. It is noteworthy that the slope of the bulk $A-B$ phase transition line actually changes its sign around the PCP, with a positive slope for $P<P_c$. The observed behavior of the strong-coupling parameter, $g(\beta)$, and these differences can be accounted for qualitatively and naturally by recognizing the reduction in strong-coupling effects due to impurity-scattering.\cite{thunne,schoi,aoyama,aoyama07} Briefly and simply stated, these effects combine to effectively shift the phases and features of the bulk phase diagram up in the pressure to yield the phase diagram for $^3$He in 98\% aerogel.
\begin{figure}
\includegraphics[width=3.7in]{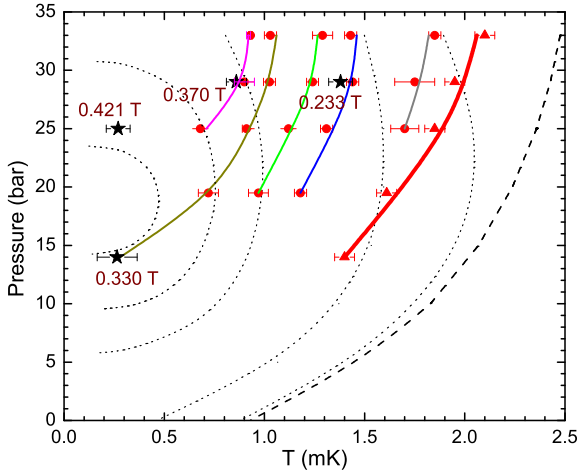}
\caption{\label{Fig.13.} (Color online) Phase diagram of superfluid $^3$He in 98\% aerogel.  The solid triangles represent the aerogel superfluid transition. The $A-B$ transitions in aerogel obtained by the TSCF are in solid circles and by the IFS in solid stars. The solid lines going through the data points are guide for eyes but conforms to the constant field phase boundaries for 0.111, 0.222, 0.275, 0.333, and 0.385~T, respectively from right to left. For comparison, the constant field $A-B$ phase boundaries for the bulk liquid are shown by the dotted lines\cite{hahn} for 0.1, 0.3, 0.5, 0.55, and 0.58~T, respectively. The numbers right next to the star symbols indicate the mid-field strength of the transition.}
\end{figure}

In G-L theory, the free energy (relative to the normal state) of the $A(B)$ phase is $f_{A(B)} = -\alpha^{2}/{2\beta_{A(B)}}$, where $\alpha = N(0)(T/T_{c}-1)$ is the coefficient of the quadratic term in the G-L free-energy expansion, $N(0)$ is the density of states at the Fermi surface, and $\beta_{A}=\beta_{245}$, $\beta_{B}=\beta_{12}+\beta_{345}/3$.   In zero field, the two phases share the same superfluid transition temperature and the PCP is determined by the condition $\beta_{A}(P_c)=\beta_{B}(P_c)$.  The presence of a magnetic field introduces an additional term in the G-L expansion given by
\begin{equation}
f_z = g_{z}B_{\mu}A_{\mu i}A_{\nu i}^*B_{\nu} \label{1}.
\end{equation}
Here, $A_{\mu i}$ represents the order parameter of a superfluid state with spin ($\mu$) and orbital ($i$) indices.\cite{vollhardt}  The magnetic field couples through the spin channel of the order parameter. With two distinct symmetries in the $A$ and $B$ phase order parameters, this quadratic contribution lifts the degeneracy in the superfluid transition temperature, thereby pushing the $A$-phase $T_{c}$ slightly above that of the $B$ phase.  As a result, a narrow region of the $A$-phase must be wedged between the normal and the $B$ phase for $P<P_c$, even for an infinitesimally small magnetic field.  The degree of this effect is inversely related to the free-energy difference between two phases, $g(\beta)\propto (\beta_{A}-\beta_{B})^{-1}$, giving rise to the diverging behavior in $g(\beta)$ as $P\rightarrow P_c$.

In the presence of aerogel, the impurity-scattering warrants various corrections to both $\alpha$ and $\beta$ parameters.  The first-order corrections obviously come from the suppression of $T_c$ by pair-breaking and incur the reduction in the strong-coupling effects in the $\beta$-parameters simply scaled by $T_{ca}/T_c$.  The most extensive calculation of the $\beta$-parameters including various vertex corrections was done by Aoyama and Ikeda.\cite{aoyama07}  Their theoretical phase diagram based on those corrections indeed resembles the bulk phase diagram that is, in effect, shifted to lower temperature and, simultaneously, to higher pressure, resulting in the relocation of the PCP to a higher pressure.

Aoyama and Ikeda have also incorporated the anisotropic nature of the aerogel through the angular dependence of the scattering amplitude.\cite{aoyama}  In a uniaxially deformed aerogel, the calculation shows the unambiguous effect of global anisotropy as uniform orbital field, represented by an additional quadratic free-energy term,\cite{thunne}
\begin{equation}
f_a = g_{a}a_{i}A_{\mu i}A_{\mu j}^{*}a_{j} \label{2},
\end{equation}
where $\hat{a}$ is a unit vector pointing in the direction of the aerogel strand.  The similarity between Eqs.\,(5) and (6) is apparent.  The effect of the orbital field produced by the aerogel strands was estimated to be comparable to the effect produced by a magnetic field $\sim$~0.1~T in the case of complete alignment.\cite{vicente}  It has been experimentally demonstrated that uniaxial compression indeed induces optical birefringence proportional to the strain and, consequently, global anisotropy into the system.\cite{pollanen,bhupathi}

In a globally isotropic aerogel, however, the local anisotropy  comes into play only when $\xi\lesssim\xi_{a}$, where $\xi_{a}$ represents the correlation length of the aerogel and $\xi$ is the pair correlation length.\cite{vicente} In the other limit, the local anisotropy is simply averaged out to produce no effect.  As discussed by Vicente {\it et al.\,}, this net local anisotropy should emulate the effect of magnetic field even in the absence of magnetic field in a globally isotropic aerogel.  Furthermore, an inhomogeneity in the local anisotropy would cause a broadening of the $A-B$ transition in aerogel in which the mixture of the $A$ and $B$ phases coexists.\cite{imry}  Considering $\xi_{a}\approx$~40~-~50~nm in 98\% aerogel, this local anisotropy effect in a globally isotropic aerogel should be more pronounced at higher pressures but is expected to tail off as the pressure decreases to the point where $\xi \sim \xi_{a}$, which occurs around 10~bar.  The impressive agreement in $T_{ca}$ between the experiments and the theory of Sauls and Sharma\cite{sauls03} was achieved by incorporating the aerogel correlation length into the depairing parameter of the homogeneous isotropic scattering model.\cite{thunne}

Although the aerogel sample used in this work is supposed to be isotropic, we cannot rule out the possibility of having a weak global anisotropy built into this sample from the sample preparation or the shrinkage occurring during condensation of $^3$He.  In either case, the observed behavior in this work as well as others can be explained coherently.\cite{herman,pollanen}


\section{Conclusion}
Longitudinal ultrasound attenuation measurements were conducted in a 98\% uncompressed aerogel in the presence of magnetic fields.  Utilizing the metastable $A$-like phase that extended down to the lowest temperature in 0.444~T, we were able to establish the temperature dependence of the attenuation in the $A$-like phase over the entire superfluid region.  This arrangement allowed us to determine the $A-B$ transitions in aerogel in various magnetic fields.  Based on the transition points on warming, a $P$-$T$-$B$ phase diagram of this system is constructed.  The key features of the phase diagram can be understood on the basis of two fundamental points: first, the strong-coupling effect is significantly reduced in this system by impurity-scattering, and second, the anisotropic disorder presented in the form of aerogel strands plays an important role that emulates the effect of a magnetic field.

\begin{acknowledgments}
We wish to acknowledge the support of the NSF under Grants No. DMR-0803516 (Y.L.), No. DMR-0701400 (M.W.M.), No. DMR-0654118 (NHMFL), and the State of Florida.
\end{acknowledgments}

\newpage 

\begin{thebibliography}{99}
\bibitem{note1} The same sweep rate was used for all isothermal field sweeps performed in this work.
\end{thebibliography}

\end{document}